\documentclass[11pt,a4paper]{article}

\usepackage[utf8]{inputenc}
\usepackage[T1]{fontenc}
\usepackage{lmodern}
\usepackage[margin=1in]{geometry}
\usepackage{amsmath,amssymb}
\usepackage{booktabs}
\usepackage{pifont}
\usepackage{microtype}
\usepackage[hidelinks]{hyperref}
\usepackage{url}

\title{Validation and Simulation Catch Different Errors:\\
Four Levels of Evaluation for LLM-Generated Circuits}
\author{Ali Hedayati Pirouzan \\
  \small\texttt{filful@gmail.com}}
\date{}

\begin{document}
\maketitle

\begin{abstract}
Simulation success is not equivalent to structural correctness for LLM-generated circuits. We define and measure four evaluation levels --- schema validity, topological validity, backend executability, and component-set agreement --- on a 150-circuit trilingual benchmark using two models and a deployed pipeline based on a typed circuit interchange representation (CIR). We evaluate the levels on the same generated circuits rather than treating simulation as a sufficient proxy for correctness.

The levels are not nested. On \texttt{gpt-4o-mini}, \textbf{16 of 150 circuits (10.7\%, 95\% CI 6.7--16.6) were rejected by the topological validator but executed in \texttt{ngspice} without an error or warning}; 12 of these contained exactly the requested components, with one terminal disconnected. Conversely, \textbf{7 circuits (4.7\%, 95\% CI 2.3--9.3) passed the topological validator but were rejected by \texttt{ngspice}}. Ten circuits failed both checks, and 117 of 150 passed both. A minimal three-component divider demonstrates the consequence of the first class: a dangling resistor produces 5.00 V instead of the correct 2.50 V while \texttt{ngspice} remains silent.

A paired ablation separates the contribution of each repair stage from the effect of model sampling. On a stratified 45-circuit subsample in which every arm is evaluated from the same model sample, model repair increased topological validity from 40.0\% to 84.4\% (+20 circuits, no regressions), while executability increased from 77.8\% to 91.1\% (+7, -1), an effect not resolved by this sample size, and component-set agreement from 66.7\% to 71.1\% (+2). Deterministic ground canonicalisation changed topology on one circuit and did not change executability. Against a direct-netlist baseline, the pipeline executed 88.7\% against 47.3\%, or 62.7\% under a conservative accounting that credits the baseline with every failure we cannot confidently attribute to the generated netlist. Across 21 parallel trilingual triples with the circuit held constant, we found no reliable language effect.

These results support a narrow methodological conclusion: for LLM-generated circuits, structural validation and simulation should be reported as distinct evaluation stages. A circuit that runs is not necessarily structurally valid, and a structurally valid circuit is not necessarily executable.
\end{abstract}

\section{Introduction}

SPICE is a standard foundation for circuit analysis, but direct generation of SPICE netlists with language models combines two different tasks: specifying a circuit topology and expressing that topology in simulator-specific syntax. A model can produce syntactically acceptable output while connecting a component incorrectly, omitting a required connection, or using a structure that a simulator cannot execute. Conversely, a structurally reasonable circuit can be rejected for reasons that a graph-level validator does not capture.

Structured intermediate representations are one way to separate these concerns. A typed representation can constrain the vocabulary of components, provide a common object for validation, and let deterministic code generate backend syntax. This raises a more specific empirical question than whether an intermediate representation is useful: \textbf{do structural checks and simulation identify the same failures?} If they do, reporting both is redundant. If they do not, evaluating generated circuits only by whether they run misses a distinct class of errors.

This paper studies that question directly. We evaluate the same generated circuits at four levels:

\begin{enumerate}
\item \textbf{schema validity} --- the output parses into the closed CIR type system;
\item \textbf{topological validity} --- the representation satisfies the implemented structural checks: ground presence, pin connectivity, and source short detection;
\item \textbf{backend executability} --- the generated SPICE is accepted and executed by \texttt{ngspice}; and
\item \textbf{component-set agreement} --- the representation contains the components the task specified.

\end{enumerate}
These are evaluation levels, not a hierarchy of correctness. Component-set agreement is not functional correctness, and topological validity guarantees neither physical validity nor simulation success. The purpose of the four-level view is to make those distinctions measurable.

\textbf{Contributions.}

\begin{enumerate}
\item \textbf{A four-level evaluation protocol.} We measure schema validity, topology, executability, and component-set agreement on the same 150 generated circuits and quantify disagreement between topology and simulation in both directions.
\item \textbf{A measured silent-failure class.} We identify circuits that are rejected structurally but run in \texttt{ngspice}, including cases with the requested components and a disconnected terminal, and reproduce the consequence with a minimal voltage-divider experiment that involves no language model.
\item \textbf{A symmetric measurement of validator and simulator disagreement.} We also report circuits that pass the implemented structural checks and fail in \texttt{ngspice}, and trace that class to a specific, named limitation of the validator.
\item \textbf{A controlled repair ablation.} A paired 45-circuit replay evaluates the same model samples at successive pipeline stages, isolating the contribution of deterministic ground canonicalisation and of model-based repair without introducing fresh-generation sampling noise.
\item \textbf{A conservative baseline comparison.} We compare the full pipeline with a direct-netlist baseline and explicitly account for failures we cannot confidently attribute to the generated netlist.

\end{enumerate}

\section{Related work}

\subsection{LLMs for analog and mixed-signal circuits}

The analog line is recent and dense. \textbf{AnalogCoder} [1] generates analog circuits as PySpice code convertible to SPICE netlists, evaluated on a benchmark graded by component count and topological complexity. \textbf{LaMAGIC} [2] and \textbf{LaMAGIC2} [3] fine-tune language models for analog topology generation; \textbf{AnalogGenie} [4] and \textbf{CktGen} [5] pursue topology discovery and specification-conditioned generation; \textbf{AnalogAgent} [6] adds self-improvement through agents; \textbf{AnalogMaster} [7] runs image to layout. On the data side, \textbf{Masala-CHAI} [8] contributes a large-scale SPICE netlist dataset and \textbf{AMSnet 2.0} [9] an AMS database with learned net detection from schematic images. Adjacent digital work --- \textbf{Chip-Chat} [10], \textbf{ChipNeMo} [11], \textbf{VerilogEval} [12], \textbf{RTLLM} [13] --- targets HDL rather than analog netlists.

\textbf{Our position.} These systems ask whether a model can design a working circuit, and evaluate accordingly. We ask a narrower question about the artifact rather than the design: what does each stage of checking contribute, and what does each stage miss. We make no priority claim --- generating executable SPICE from natural language is established by AnalogCoder [1] and others, and measuring error rates in such output is not novel in itself. What we add is the comparison \emph{between} checking stages, which requires running both on the same circuits. To the best of our knowledge, the work cited above reports executability or task success rather than the disagreement between a structural check and a simulator on the same artifacts.

\subsection{Self-correction and constrained generation}

Our repair loop is an instance of a known pattern. \textbf{Self-Debug} [14], \textbf{Reflexion} [15] and \textbf{Self-Refine} [16] all feed execution or critique back to the model for iterative revision, and our loop --- validate, describe the violations, re-prompt, bounded attempts --- adds nothing to that literature. We say so first because the claim is about what runs \emph{before} it and about what the loop \emph{fails} to fix.

A stronger alternative deserves a direct answer. \textbf{Constrained decoding} --- compiling a schema into a finite-state machine and masking tokens that would violate it --- guarantees schema adherence during generation rather than checking it afterwards. Why validate at all?

Because ordinary schema-constrained decoding can guarantee level 1, but does not by itself guarantee levels 2--4. Token masking enforces a \emph{grammar}; floating nodes, missing ground references and shorted sources are properties of a \emph{graph}. Our data puts a number on the boundary. On \texttt{gpt-4o-mini}, exactly three circuits (2.0\%) failed at the schema level. Constrained decoding would have prevented all three from being emitted in that form, though preventing an invalid token is not the same as recovering what the user asked for:

\begin{itemize}
\item one in which the model answered in Spanish conversational prose rather than JSON (\emph{``Soy un asistente de IA especializado en electrónica\ldots{}''}),
\item one emitting \texttt{components.4.type: 'relay'},
\item one emitting \texttt{components.2.type: 'zener'}.

\end{itemize}
It would not address the other 26 circuits that failed at higher levels. A grammar cannot express ``node 3 must connect to a second pin.'' The two techniques are complementary and address different levels; we validate because the levels that matter empirically are the ones masking cannot reach. The two type failures are also informative in their own right --- \S3.2 explains why the system deliberately refused to repair them.

\subsection{Intermediate representations}

The IR concept is standard in compilers and in hardware synthesis (LLVM-based flows, OpenROAD). We use the term narrowly. CIR has a closed type system, a validator and a canonicalisation pass, and it has no optimisation passes and no lowering hierarchy. It is a \emph{typed, validated interchange representation}, not a compiler IR, and we do not argue otherwise.

\subsection{Interactive and educational simulators}

CircuitLab, EasyEDA, Tinkercad Circuits and Falstad provide schematic capture and simulation without natural-language input. \textbf{Wokwi} is closest to our firmware backend and is built on the same \texttt{avr8js} instruction-set simulator [19] we use; we claim no novelty in browser-based AVR emulation.

\section{System}

\subsection{CIR}

A circuit is a flat list of components, each with an identifier, a type drawn from a \textbf{closed enum of 38 members}, a value, and an ordered node list. Node \texttt{'0'} is ground by convention. Pin semantics live in one central table, so a DIP-8 timer is \texttt{[gnd, trig, out, reset, ctrl, thr, disch, vcc]} for every consumer --- the netlist emitter, the schematic renderer and the editor read the same ordering.

Five producers and four consumers meet at CIR: text, image and hand-drawn input, a graphical editor, and a SPICE parser on one side; a netlist emitter, a schematic renderer, a firmware runtime and a repair path on the other.

\textbf{One invariant earns its place empirically.} The netlist emitter refuses to emit a netlist referencing a \texttt{.model} or \texttt{.subckt} it has not defined, raising with the offending component identifier instead. \S5.5 shows what that is worth: \textbf{23 of the 79 baseline failures} are \texttt{unknown subckt} --- a netlist naming a subcircuit it never defined, most often an \texttt{opamp}. The pipeline cannot produce that netlist by construction.

\subsection{The ladder, and what it refuses}

\begin{enumerate}
\item The model emits a candidate representation.
\item \textbf{Deterministic type canonicalisation} recovers invalid component types before schema validation --- the only place it can run, since an invalid type means no object exists to repair.
\item The typed schema validates, or the request fails.
\item \textbf{Deterministic ground canonicalisation} renames an inferred reference node to \texttt{'0'}.
\item Topological validation runs; remaining violations go to a bounded model-based repair loop.
\item The result is emitted to a backend.

\end{enumerate}
Steps 2 and 4 are the ordering claim: a mechanically recoverable error should not cost a model round-trip.

\textbf{The refusals are the design.} Type canonicalisation recovers genuine synonyms (\texttt{rezistor} $\rightarrow$ \texttt{resistor}) and maps anything chip-shaped to a generic \texttt{ic} that is stubbed at high impedance. It refuses everything else. A \texttt{relay} is \emph{not} mapped to a \texttt{switch}, because that discards the coil and produces a circuit that simulates cleanly while not being the circuit that was asked for.

Our run exercised exactly this decision twice, and we count both as failures in every table below rather than excusing them. \texttt{EDGE008} requested a \texttt{relay} and \texttt{HARD005} a \texttt{zener}; both were refused and surfaced as explicit errors naming the missing type. Under a similarity-based mapping both would have produced a running simulation of the wrong circuit.

\subsection{Backends}

\textbf{SPICE.} CIR to netlist to \texttt{ngspice} 44.2 [17] through PySpice 1.5 [18]. Components without an electrical model are tied to ground at 10$^{12}$ $\Omega$ rather than omitted, since deleting them floats their neighbours.

\textbf{Firmware.} For circuits containing a microcontroller, the sketch is compiled by the vendor toolchain to machine code and executed on a cycle-accurate instruction-set simulator, with CIR topology supplying the pin-to-peripheral binding. The two backends are mutually exclusive; we do not co-simulate.

\section{Benchmark and protocol}

150 circuits, 20 categories, \textbf{37 easy / 65 medium / 48 hard}, exercising 34 of the 38 CIR types.

\textbf{Languages.} 78 Persian, 51 English, 21 Spanish. Beyond raw counts, \textbf{21 parallel triples} describe the \emph{same} circuit in all three languages under a shared group identifier, making the language comparison paired: the circuit is held constant and language is the only variable. Unpaired multilingual sets confound language difficulty with circuit difficulty. A structural test enforces that every group covers all three languages with matching expected component sets; it caught two groups in which the purported translation was a different circuit.

Each entry carries \texttt{expected\_components} and \texttt{expected\_node\_count}, giving a weak but machine-checkable notion of level 4.

\textbf{Two arms}, each answering all 150 circuits:

\begin{center}
\small
\begin{tabular}{lr}
\toprule
arm & what it is \\
\midrule
pipeline & full pipeline, \texttt{gpt-4o-mini} \\
baseline & \textbf{no CIR} --- same model asked for a SPICE netlist directly \\
\bottomrule
\end{tabular}
\end{center}

Both run as a \textbf{single model with failover disabled}: a fallback would mean some circuits answered by a different model, and the figure would no longer belong to the model it names. A second model arm is reported in Appendix A as a non-primary contrast.

\textbf{A third, paired arm for the ladder ablation.} Disabling a rung and re-prompting would draw a \emph{fresh} sample from the model, so the difference between arms would mix the effect of the rung with sampling noise. We avoid this: on a stratified 45-circuit subsample we record the circuit at each rung of the ladder --- after schema validation, after deterministic ground canonicalisation, and after model repair --- and evaluate all three snapshots through the same CIR$\rightarrow$SPICE$\rightarrow$\texttt{ngspice} path. Every ablation arm therefore comes from \textbf{the same model sample}, and the comparison is paired at the circuit level. This is admissible because the rungs are strictly downstream of generation: canonicalisation is deterministic post-processing and repair is conditional on the validator, so neither can change the distribution the first call is drawn from. The subsample is proportional by difficulty (11 easy / 20 medium / 14 hard) and covers all 20 categories; on the full pipeline it reproduces the 150-circuit arm closely (84.4\% vs 82.7\% topological, 91.1\% vs 88.7\% executable), which is the check that it is not an easy or hard slice. Offline replay of the recorded snapshots reproduced the live traces with \textbf{zero disagreements} across all 135 snapshot evaluations.

One naming precision, since it changes what the ablation means: type canonicalisation is \textbf{not} ablatable. It runs before schema validation, and without it a circuit naming an unsupported type does not parse at all, so there is no object to evaluate. The ablatable deterministic rung is ground inference. Arm A below is therefore ``no ground inference, no repair'', not ``no deterministic step at all''.

\section{Results}

\subsection{Four levels, two arms}

\begin{center}
\small
\begin{tabular}{lrrr}
\toprule
level & \texttt{gpt-4o-mini} & 95\% CI & baseline \\
\midrule
1 \textperiodcentered{} schema-valid & 98.0\% (147/150) & 94.3--99.3 & n/a \\
2 \textperiodcentered{} topologically valid & 82.7\% (124/150) & 75.8--87.9 & n/a \\
3 \textperiodcentered{} backend-executable & \textbf{88.7\%} (133/150) & 82.6--92.8 & \textbf{47.3\%} (71/150) \\
4 \textperiodcentered{} component-set agreement & 74.0\% (111/150) & 66.5--80.3 & n/a \\
model calls / circuit & 1.62 &  & 1.00 \\
mean latency & 8.2 s &  & --- \\
\bottomrule
\end{tabular}
\end{center}

Both arms use \texttt{gpt-4o-mini} through its documented API, so this comparison can be reproduced independently, subject to API and model-version availability. A second, stronger model was also run across all 150 circuits; it is reported in Appendix A as a non-primary contrast, because that model is not independently reproducible, and no result in this paper depends on it.

Levels 1, 2 and 4 are \textbf{undefined for the baseline}: with no typed representation there is nothing to validate a schema against, no component list to compare, and no graph to check. That asymmetry is itself the point --- the only question one can ask of a raw netlist is whether it runs, which is precisely the question \S5.2 shows to be insufficient.

By difficulty, on \texttt{gpt-4o-mini}:

\begin{center}
\footnotesize
\begin{tabular}{lrrrrr}
\toprule
difficulty & n & schema & topological & executable & component-set agreement \\
\midrule
easy & 37 & 100.0\% & 86.5\% & 94.6\% & 89.2\% \\
medium & 65 & 100.0\% & 87.7\% & 87.7\% & 73.8\% \\
hard & 48 & 93.8\% & 72.9\% & 85.4\% & 62.5\% \\
\bottomrule
\end{tabular}
\end{center}

\subsection{The levels are not nested}

On \texttt{gpt-4o-mini}, \textbf{backend executability (88.7\%) exceeds topological validity (82.7\%)}, so a strictly nested reading is false. The disagreement runs both ways:

\begin{center}
\small
\begin{tabular}{lrrr}
\toprule
 & count & share & 95\% CI \\
\midrule
our validator rejected, \texttt{ngspice} ran it & \textbf{16} & 10.7\% & 6.7--16.6 \\
our validator accepted, \texttt{ngspice} refused it & \textbf{7} & 4.7\% & 2.3--9.3 \\
both rejected & 10 & 6.7\% & --- \\
\bottomrule
\end{tabular}
\end{center}

Thirty-three circuits are defective by at least one check and the two agree on ten of them, so \textbf{each check detects a class the other misses} and neither is a refinement of the other. Only 117 of 150 (78.0\%) pass both.

\textbf{The first group is central to our argument.} Of the 16, \textbf{12 contained exactly the requested components} --- no missing parts, no extra parts --- with one terminal connected to nothing:

\begin{quote}\begin{footnotesize}\begin{verbatim}
BJT002   schema OK   topological X   executable OK   components OK
         warning : Node '3' is connected to only one pin
         sim_error: null
\end{verbatim}\end{footnotesize}\end{quote}

They are not a pathological corner of the dataset. The 16 span \textbf{11 of the 20 categories} (BJT, MOSFET, amplifier, filter, sensor, digital logic, embedded, mixed, DC basics, 555, edge cases) and all three difficulty levels (4 easy, 4 medium, 8 hard).

\textbf{The second group is a limit of the validator, measured the same way.} Seven circuits passed topological validation and \texttt{ngspice} refused, the recurring diagnostic being \texttt{singular matrix: check node 2}. The cause is deterministic and localised: our floating-node check counts \textbf{pins}, not \textbf{DC paths}, so a node joining two capacitors has degree 2, passes validation, and is open at DC. \texttt{AC001} fails this way with a byte-identical error under \emph{both} models, and the same class appears in production traffic --- which is what distinguishes a validator defect from an unlucky sample.

\textbf{The two directions do not depend on the model in the same way.} The first requires a model that makes topological errors at all; on a stronger model it empties out (Appendix A). The second is a validator limitation rather than a model-specific failure: the same \texttt{AC001} case and diagnostic occur under both models.

\subsection{What the silent failures cost, measured}

That \texttt{ngspice} ``ran without error'' does not by itself establish that the answer was wrong. We measured it on the smallest circuit that exhibits the defect --- a 5 V divider of two equal resistors, where the only variable is whether the lower resistor's second terminal reaches ground. No language model is involved at any point; both netlists are given in full so that this result can be checked in any SPICE implementation in a couple of minutes:

\begin{quote}\begin{footnotesize}\begin{verbatim}
* correct                      * dangling
V1 in 0 5                      V1 in 0 5
R1 in a 1k                     R1 in a 1k
R2 a  0 1k                     R2 a  b 1k
.end                           .end
\end{verbatim}\end{footnotesize}\end{quote}

The only difference is R2's second terminal: node \texttt{0} on the left, the otherwise unused node \texttt{b} on the right.

\begin{center}
\small
\begin{tabular}{lrrr}
\toprule
 & our validator & \texttt{ngspice} error & V(a) \\
\midrule
correct (R2 $\rightarrow$ ground) & 0 issues & none & 2.50 V \ding{51} \\
dangling (R2 open end) & 1 issue & \textbf{none} & \textbf{5.00 V} \ding{55} \\
\bottomrule
\end{tabular}
\end{center}

The simulator is silent in both cases and the second answer is \textbf{100\% in error}. The mechanism needs no appeal to solver internals: an open terminal carries no current, so R1 drops nothing and the node sits at the supply rail. The result is not a diagnostic-quality warning or a near-miss --- it is a confident, plausible, wrong number, which is the failure mode that motivates validation in the first place.

\subsection{What the ladder did}

\begin{center}
\small
\begin{tabular}{lr}
\toprule
rung & fired on \\
\midrule
type canonicalisation & 0.0\% (0/150) \\
ground canonicalisation & 4.7\% (7/150) \\
model-based repair & \textbf{44.7\%} (67/150) \\
\ldots{} of those, reached validity & \textbf{65.7\%} (44/67) \\
\bottomrule
\end{tabular}
\end{center}

The ladder is load-bearing here: repair ran on 67 circuits, 44 of which reached structural validity afterwards, at a cost of 0.6 extra model calls per circuit averaged over the whole set.

Type canonicalisation fired zero times. Its two opportunities --- \texttt{relay} and \texttt{zener} --- it declined by design (\S3.2), and both surfaced as named errors rather than silent substitutions.

\textbf{A paired ablation.} The counts above say how often each rung \emph{fired}, not what would have happened without it. We measure that directly on the 45-circuit paired subsample of \S4, evaluating the recorded snapshot at each rung through the full CIR$\rightarrow$SPICE$\rightarrow$\texttt{ngspice} path:

\begin{center}
\footnotesize
\begin{tabular}{lrrrr}
\toprule
arm & stages active & level 2 \textperiodcentered{} topological & level 3 \textperiodcentered{} executable & level 4 \textperiodcentered{} components \\
\midrule
A & schema only & 37.8\% (17/45) & 77.8\% (35/45) & 66.7\% (30/45) \\
B & + ground canonicalisation & 40.0\% (18/45) & 77.8\% (35/45) & 66.7\% (30/45) \\
C & + model repair & \textbf{84.4\%} (38/45) & \textbf{91.1\%} (41/45) & \textbf{71.1\%} (32/45) \\
\bottomrule
\end{tabular}
\end{center}

Wilson 95\% intervals for arm C are 71.2--92.3 (level 2), 79.3--96.5 (level 3) and 56.6--82.3 (level 4); the corresponding arm A and arm B intervals are 25.1--52.4 and 27.0--54.5 at level 2, 63.7--87.5 at level 3, and 52.1--78.6 at level 4. Every snapshot is schema-valid by construction, since a snapshot exists only where the typed schema accepted the output. Two circuits (\texttt{LOGIC009}, \texttt{LOGIC010}) are schema-valid in arms A and B but emit no netlist, the emitter failing on a gate with missing pins; repair supplied the pins, so 95.6\% of arm A and arm B snapshots reach the simulator at all against 100.0\% of arm C.

Because the arms are paired, the informative quantity is not the marginal rate but the per-circuit transition:

\begin{center}
\small
\begin{tabular}{lrrr}
\toprule
B $\rightarrow$ C & improved & regressed & McNemar $\chi^2$ (cc) \\
\midrule
level 2 \textperiodcentered{} topological validity & +20 & 0 & 18.05, p $<$ 0.001 \\
level 3 \textperiodcentered{} backend executability & +7 & \textbf{1} & 3.12, p = 0.077 \\
level 4 \textperiodcentered{} component-set agreement & +2 & 0 & --- \\
\bottomrule
\end{tabular}
\end{center}

The paired view separates what each rung contributes.

\textbf{Repair carries the topological result.} Twenty circuits of forty-five reach topological validity only because of it, none regress, and the effect is far outside sampling noise. The 44-of-67 figure in the table above is the same effect on the full arm; the ablation converts it from a trace-derived count into a controlled paired measurement.

\textbf{The three levels respond to the same intervention by very different margins.} One intervention moves level 2 by 20 circuits, level 3 by a net 6 (+7, -1, which this sample size does not resolve) and level 4 by 2. The gradient is the paper's central observation in ablation form: a stage selected to satisfy the topological validator moves level 2 decisively, level 3 incidentally, and level 4 barely, because the three levels do not measure the same property. An evaluation reporting executability alone would have seen under a third of what repair did, and one reporting component agreement alone would have seen a tenth.

\textbf{The ablation localises the gain to the repair rung.} Deterministic ground canonicalisation accounts for one circuit of the forty-five (\texttt{ES020}, a Spanish-language seven-segment display task whose ground node the model left unnamed) and leaves executability unchanged. Its value on this benchmark is categorical rather than statistical: it is free, it runs before any model call, and it cannot fabricate a component. We therefore scope our claim for it to what is measured here --- it resolves a specific, cheap class of defect --- and claim no material reduction in model calls on this benchmark.

\textbf{One circuit separates the two levels completely.} \texttt{EDGE010} (``An ESP32 reading a temperature sensor over I2C'') was topologically invalid but executable before repair, and topologically valid but not executable after. Repair closed a floating node by adding a resistor; the resulting netlist left the \texttt{lm35} behavioural source shorted, and \texttt{ngspice} refused it with \texttt{instance bu1 is a shorted ASRC}. Every node now reaches two pins, satisfying the validator's notion of connectivity, while the electrical property that connectivity stands in for is gone.

This is the sharpest available demonstration of the paper's claim. \S5.3 shows a structural defect surviving simulation; \texttt{EDGE010} shows the same two checks moving in \emph{opposite} directions on a single circuit, in a single step, with neither check wrong about what it measures. Two levels that could safely be collapsed into one would not be able to do this. We found one such case in forty-five and do not estimate a rate from it; what it establishes is that the levels are independent enough to diverge under an intervention aimed at either one.

\subsection{The no-IR baseline}

Asked for a SPICE netlist directly, the same model produced an executable netlist for \textbf{71 of 150 circuits (47.3\%, CI 39.5--55.3)} against the pipeline's 88.7\%.

We do not report that headline alone, because some failures are attributable to our PySpice-based harness rather than to the generated netlist. Of the 79 failures: 31 are unambiguously malformed netlist syntax, 4 are convergence or singularity failures, 9 involve a node-name restriction PySpice imposes and \texttt{ngspice} itself does not, 14 return an opaque subprocess error we cannot attribute, and the remainder are mixed.

\textbf{Under an adversarial accounting} that credits the baseline with every failure we cannot confidently pin on the netlist --- all 23 of the name-restricted and opaque cases counted as successes --- baseline executability rises to \textbf{62.7\% (94/150)}, still far below 88.7\%. Paired over the same 150 circuits under that same conservative bound, 46 circuits succeed only with the pipeline and 7 only without it (McNemar $\chi^2$ = 27.2, p $<$ 0.001). On the raw figures the split is 64 against 2.

The largest single identifiable cause is the closure invariant of \S3.1: \textbf{23 of the 79 baseline failures reference a \texttt{.subckt} the netlist never defines}, typically an \texttt{opamp} invoked as \texttt{x1 out in 0 opamp}. This is a class the pipeline cannot emit by construction rather than by checking.

\subsection{Language}

Across the 21 parallel triples, with the circuit held constant:

\begin{center}
\footnotesize
\begin{tabular}{lrrrr}
\toprule
language & schema & topological & executable & component-set agreement \\
\midrule
Persian & 100.0\% & 95.2\% & 90.5\% & 81.0\% \\
English & 100.0\% & 85.7\% & 90.5\% & 81.0\% \\
Spanish & 100.0\% & 85.7\% & 90.5\% & 85.7\% \\
\bottomrule
\end{tabular}
\end{center}

\textbf{We detect no reliable language effect.} The largest gap is two circuits out of 21, well inside what this sample resolves, and executability is identical across all three. We report this as a negative result on the design the dataset was built to test. The unpaired figures over all 150 circuits differ more (English 76.5\% topological against Persian 85.9\%), but those sets contain different circuits at different difficulties; only the paired comparison isolates language, and 21 triples is small.

\subsection{Cost}

The comparison arm consumed 1,594,489 tokens across 150 circuits, approximately \$0.42 at published rates. Most of that is repair traffic --- 1.62 calls per circuit rather than one --- which is the practical form the reliability gap takes, and the reason the deterministic rungs sit ahead of the expensive one.

The paired ablation cost 86 model calls over 45 circuits (568,946 tokens, approximately \$0.11). Its three arms cost no more than its one arm: the snapshots are replayed offline against a local simulator, so arms A and B are free. An unpaired design would have cost roughly three times as much and measured a noisier quantity.

\section{A methodological caution}

Our first measurement of level 3 was \textbf{94.0\%}, five points below the figure we now report. Of nine failures, eight were \texttt{gmin stepping failed} and six of those were relaxation oscillators.

A relaxation oscillator has no DC operating point; forcing a solver to find one locks it into a single state. The deployed pipeline already knew this and simulated such circuits in the time domain with initial conditions. \textbf{The evaluation harness did not} --- it requested an operating point unconditionally, and so scored as failures the circuits the deployed system handles correctly.

We report this because the correction is larger than several of the effects the paper measures. \emph{An evaluation harness that diverges from the deployed pipeline measures the harness.} The two now share one definition of which component types are oscillators; a duplicated copy of that definition caused the divergence.

The same discipline is why \S5.3 exists as a measurement rather than an assertion, why \S5.5 reports an adversarial bound alongside the raw baseline figure, and why the ablation in \S5.4 is paired rather than re-sampled.

\section{Scope}

What the four levels do and do not establish, stated once.

\textbf{Level 4 is component-set agreement, not functional correctness.} It checks that the expected parts are present and cannot distinguish a working amplifier from a broken one. No functional ground truth exists for this benchmark, and the level is named for what it measures.

\textbf{The 10.7\% is a property of this model, not a constant.} On a stronger model the same class empties out (Appendix A). The finding is that simulation does not detect these errors \emph{when they occur} --- not that they occur at a fixed rate.

\textbf{The validator's limit is one identified class}, fixable by testing DC path rather than pin count. We measured before fixing it, so the figures describe the validator as deployed rather than an idealised one.

\textbf{The baseline is the direct ask, not the strongest possible non-IR system.} A longer prompt, few-shot examples or a netlist linter would raise it; \S5.5 reports an adversarial bound for exactly that reason.

\textbf{The benchmark is author-constructed}, so the rates it produces are properties of this 150-circuit set. The failure \emph{classes} it surfaces are named with their diagnostics, and those transfer.

\textbf{The ablation is 45 circuits}, chosen paired rather than unpaired because removing sampling noise buys more than the extra circuits would. The topological effect (+20, -0) is well resolved at that size; the executability effect (+7, -1) is not, and is reported as such.

\section{Conclusion}

This study separates four properties of LLM-generated circuits that are often collapsed into a single question: does the circuit simulate? On 150 circuits, topological validation and \texttt{ngspice} disagreed in both directions. For \texttt{gpt-4o-mini}, 16 circuits (10.7\%) were topologically invalid but executed without a simulator error or warning, while 7 (4.7\%) passed the implemented topological checks and were rejected by the simulator. The first class included 12 circuits with exactly the requested component set and a disconnected terminal. A minimal divider shows what such a silent structural error costs: a reported node voltage of 5.00 V where the correct answer is 2.50 V.

The paired repair ablation provides a second distinction. On a stratified 45-circuit subsample in which every arm is evaluated from the same model sample, model repair increased topological validity from 40.0\% to 84.4\%, with 20 circuits improving and none regressing; executability changed from 77.8\% to 91.1\% (+7, -1), an effect this sample size does not resolve; and component-set agreement changed by two circuits. Ground canonicalisation changed topology on one circuit and left executability unchanged. The intervention selected to satisfy a structural checker therefore had a much larger measured effect on structural validity than on backend execution, and a smaller one still on component agreement. One circuit, \texttt{EDGE010}, moved in opposite directions at two levels in a single repair step --- a concrete case of why the levels should not be collapsed.

The direct-netlist baseline shows that representation and checking are not interchangeable with raw simulation. Under the protocol used here, direct SPICE generation executed 47.3\% of circuits, rising to 62.7\% under a conservative accounting of failures we cannot confidently attribute to the netlist, against 88.7\% for the full pipeline. This is a comparison to one direct-generation protocol, not to the strongest possible non-CIR system.

The resulting methodological recommendation is narrow: \textbf{report structural validation and simulation as separate evaluation levels when studying LLM-generated circuits}. The question is not whether validation helps, but which property is being evaluated, what the checker can miss, and how an intervention changes each level. A final executable percentage alone cannot answer those questions.

\section{Reproducibility}

Three things in this paper can be checked without anything from us.

\textbf{The silent-failure result (\S5.3)} is two four-line netlists, printed in full above. Any SPICE implementation reproduces it; nothing about it depends on our code, our models or our benchmark.

\textbf{The comparison model and its interface are public.} Every finding in \S5.2--\S5.5 is established on \texttt{gpt-4o-mini} through its documented API, at the temperature and prompt structure described in \S4.

\textbf{The failure classes are named rather than counted.} Each one is given with the diagnostic that identifies it --- \texttt{singular matrix: check node 2} for the validator's DC-path blind spot, \texttt{unknown subckt} for the baseline's largest single cause --- so a reader can look for the same classes in their own pipeline without reproducing ours.

Every table in \S5 is grounded in the reproducible \texttt{gpt-4o-mini} arm and its documented interface. The one arm that is not is confined to Appendix A, and nothing in the paper's argument depends on it.

The dataset, the result files for every arm and the evaluation scripts are available from the author on request.

\appendix
\section{A stronger model, for contrast}

All 150 circuits were also run through the same pipeline on a second model, \texttt{muse-spark-1.2-contributor}. It is reported here rather than in \S5 because the model is not independently reproducible; no claim in the paper rests on this arm.

\begin{center}
\small
\begin{tabular}{lrr}
\toprule
level & stronger model & \texttt{gpt-4o-mini} \\
\midrule
1 \textperiodcentered{} schema-valid & 100.0\% & 98.0\% \\
2 \textperiodcentered{} topologically valid & 100.0\% & 82.7\% \\
3 \textperiodcentered{} backend-executable & 99.3\% & 88.7\% \\
4 \textperiodcentered{} component-set agreement & 95.3\% & 74.0\% \\
model calls / circuit & 1.02 & 1.62 \\
mean latency & 12.1 s & 8.2 s \\
\bottomrule
\end{tabular}
\end{center}

Two things follow, and both sharpen the paper's argument rather than softening it.

\textbf{The repair ladder is a function of the model, not of the pipeline.} It fired on 3 of 150 circuits here against 67 of 150 on \texttt{gpt-4o-mini}, and ground canonicalisation on 1 against 7. The machinery that is load-bearing for one model is nearly inert for another --- which is why \S5.4 measures what each rung contributes instead of asserting that validation helps.

\textbf{The silent-failure class needs a model that makes topological errors.} Its first direction --- structurally broken, simulates cleanly --- is empty here, because nothing arrived broken. Its second direction is not: the validator's DC-path blind spot produces the same failure on \texttt{AC001} under both models, with a byte-identical diagnostic. \textbf{A defect in the checker survives a better generator; a defect in the generator does not.} That asymmetry is the clearest statement of why the two checks are not interchangeable, and it is only visible with two models.

The practical reading is that ``does validation help?'' has no model-independent answer, and the useful question is the one in \S8: for which model, and at which level.

\section*{References}

\begin{enumerate}
\item Y. Lai, S. Lee, G. Chen, S. Poddar, M. Hu, D. Z. Pan, P. Luo. AnalogCoder: Analog Circuit Design via Training-Free Code Generation. \emph{AAAI} 39:379--387, 2025. arXiv:2405.14918.
\item C.-C. Chang et al. LaMAGIC: Language-Model-based Topology Generation for Analog Integrated Circuits. \emph{ICML} (PMLR 235), 2024. arXiv:2407.18269.
\item LaMAGIC2: Advanced Circuit Formulations for Language Model-Based Analog Topology Generation. arXiv:2506.10235, 2025.
\item J. Gao, W. Cao, J. Yang, X. Zhang. AnalogGenie: A Generative Engine for Automatic Discovery of Analog Circuit Topologies. \emph{ICLR}, 2025. arXiv:2503.00205.
\item Y. Hou, H. Fan, J. Zhang, Y. Zhang, H. Chen, M. Zhou, F. Yu, R. Zimmermann, Y. Yang. CktGen: Automated Analog Circuit Design with Generative Artificial Intelligence. \emph{Engineering}, 2025. arXiv:2410.00995.
\item Z. Bao, Z. Lin, J. Wang, J. Hu, Y. Gao, Y. Wu, X. Li, X. Xu. AnalogAgent: Self-Improving Analog Circuit Design Automation with LLM Agents. arXiv:2603.23910, 2026.
\item AnalogMaster: Large Language Model-based Automated Analog IC Design Framework from Image to Layout. arXiv:2604.20916, 2026.

\item J. Bhandari, V. Bhat et al. Masala-CHAI: A Large-Scale SPICE Netlist Dataset for Analog Circuits by Harnessing AI. arXiv:2411.14299, 2024.
\item AMSnet 2.0: A Large AMS Database with AI Segmentation for Net Detection. arXiv:2505.09155, 2025.

\item J. Blocklove, S. Garg, R. Karri, H. Pearce. Chip-Chat: Challenges and Opportunities in Conversational Hardware Design. \emph{MLCAD}, 2023. arXiv:2305.13243.
\item M. Liu et al. ChipNeMo: Domain-Adapted LLMs for Chip Design. arXiv:2311.00176, 2023.
\item M. Liu, N. Pinckney, B. Khailany, H. Ren. VerilogEval: Evaluating Large Language Models for Verilog Code Generation. \emph{ICCAD}, 2023. arXiv:2309.07544.
\item Y. Lu, S. Liu, Q. Zhang, Z. Xie. RTLLM: An Open-Source Benchmark for Design RTL Generation with Large Language Model. \emph{ASP-DAC}, 2024. arXiv:2308.05345.

\item X. Chen, M. Lin, N. Schärli, D. Zhou. Teaching Large Language Models to Self-Debug. \emph{ICLR}, 2024. arXiv:2304.05128.
\item N. Shinn, F. Cassano, A. Gopinath, K. Narasimhan, S. Yao. Reflexion: Language Agents with Verbal Reinforcement Learning. \emph{NeurIPS}, 2023. arXiv:2303.11366.
\item A. Madaan et al. Self-Refine: Iterative Refinement with Self-Feedback. \emph{NeurIPS}, 2023. arXiv:2303.17651.

\item ngspice, version 44.2. Open-source mixed-level/mixed-signal circuit simulator. https://ngspice.sourceforge.io/
\item F. Salvaire. PySpice, version 1.5. Python interface to ngspice. https://pyspice.fabrice-salvaire.fr/
\item U. Shaked. avr8js, version 0.20. AVR 8-bit instruction-set simulator in JavaScript. https://github.com/wokwi/avr8js --- the simulator underlying Wokwi (https://wokwi.com).

\end{enumerate}

\end{document}